\documentclass[twocolumn,twoside,slac_two]{revtex4}
\usepackage{graphicx}
\usepackage{fancyhdr}
\begin{document}

\title{VERITAS Observation of M~87}

%

\author{N. Galante}
\affiliation{Harvard-Smithsonian Center for Astrophysics, 60 Garden Street, Cambridge, MA 02138 - USA}
\author{for the VERITAS Collaboration}
\affiliation{http//veritas.sao.arizona.edu}

\begin{abstract}
The giant radio galaxy M~87 is located at a distance of $\sim16$~Mpc and harbors a supermassive black hole in its center. The structure of its relativistic plasma jet is resolved at radio, optical and X-ray wavelengths. M~87 belongs to the class of active galactic nuclei (AGN) and is one of the few extragalactic TeV $\gamma$-ray source not belonging to the class of blazars. M~87 is also detected by Fermi in the GeV energy range. This makes it a unique laboratory for the study of the jet substructures and the morphology of the non-thermal emission processes. In spring 2010 a major flare was observed at TeV energies, and was sampled by VERITAS and Fermi with unprecedented accuracy. The results of the VERITAS observations will be discussed.
\end{abstract}

\maketitle

\thispagestyle{fancy}


\section{Introduction}

The search for $\gamma$-rays from radio galaxies is  important for
the understanding of the dynamics and structure of jets in active galactic nuclei (AGN).
Even though radio galaxies are AGN with jets, their jet is not oriented toward the observer
and therefore the radiation produced by the jet is not Doppler-boosted towards
higher energies and luminosities, making them more challenging to detect in the 
very high energy (VHE: $E>100$~GeV) regime.
The discovery of VHE  $\gamma$-rays from the radio galaxy M~87 by the HEGRA
collaboration~\citep{Aharonian2003}, detected later by VERITAS~\citep{Acciari2008},
and from NGC~5128 (Centaurus~A) by the HESS collaboration~\citep{Aharonian2009} has shown that
non-blazar AGN can produce very energetic photons from non-thermal processes.

Radio galaxies are classified into two main
families based on the morphology of their radio emission~\citep{FanaroffRiley},
whether it is core dominated (FR~I) or lobe dominated (FR~II),
with differences in the radio energetics and
in the discrete spectral properties~\citep{Zirbel1995}. The large number of features that
FR~I radio galaxies share with BL Lac type blazars suggests a possible unification between
the two sub-classes of AGN, in which FR I radio galaxies are BL Lac objects observed at larger jet viewing
angles~\citep{UrryPadovani}.

Evidence for synchrotron emission in radio to X-ray energies from both the extended structures and 
the core is well explained by relativistic particles moving in a beamed
relativistic jet~\citep{Ghisellini1993}. 
A commonly considered mechanism for HE-VHE (HE: high energy, 100~MeV$<E<$100~GeV) radiation is the synchrotron-self-Compton (SSC) 
process~\citep{Jones1974}, where the optical and UV synchrotron photons are up-scattered by the same 
relativistic electrons in the jet. Predictions concerning the inverse Compton (IC) component 
have long been established for the $\gamma$-ray 
emission~\citep{BloomMarscher1996} and frequency-dependent 
variability~\citep{Ghisellini1989}. Besides leptonic scenarios, several models also consider a hadronic origin for 
non-thermal emission in jets. Accelerated protons can initiate electromagnetic cascades or 
photomeson processes~\citep{Mannheim1993}, or directly emit synchrotron radiation \citep{Aharonian2002, Reimer2004}
and produce $\gamma$-rays through collisions with ambient gas \cite{Beall1999, Pohl2000}.

Modelling the blazar jet emission with a homogeneous SSC mechanism may imply particularly
high Lorentz factors, $\Gamma \gtrsim 50$, with consequent high Doppler factors and small beaming angles $\theta \simeq 1^\circ$
\citep{Kraw2002}. Such a small beaming angle is in conflict with the unification scheme according to which FR~I radio galaxies
and BL~Lac objects are the same kind of object observed at different viewing angles. Moreover,
these high values for the Doppler factor are in disagreement with the small apparent velocities observed
in the sub-parsec region of the TeV BL Lac objects Mrk~421 and Mrk~501 \citep{Marscher1999}.
These considerations suggest a more complicated geometry, for example 
a decelerating flow in the jet with a consequent gradient in the Lorentz
factor of the accelerated particles and a smaller average $\Gamma$ \citep{Georganopoulos2003}.
As a result of this gradient, the fast upstream particles interact with the downstream 
seed photons with an amplified energy density, because of the Doppler boost due to the relative Lorentz factor
$\Gamma_\mathrm{rel}$. The IC process then requires less extreme values for the
Lorentz factor and allows larger values for the beaming angle.
In a similar way, a jet spine-sheath structure consisting of a faster internal spine 
surrounded by a slower layer has been also suggested for the broadband non-thermal emission of VHE BL Lac 
objects~\citep{Ghisellini2005}. An inhomogeneous jet with a slow component may explain the HE-VHE emission observed in radio galaxies at larger angles              
($\theta_\mathrm{layer} = 1/\Gamma_\mathrm{layer} \sim 20^\circ$). 
Observation of the VHE component from radio galaxies
is therefore significant for the AGN jet modeling. In this work an overview of the observations
of radio galaxies by VERITAS is presented.

\section{The VERITAS Instrument}

The VERITAS detector is an array of four 12-m diameter imaging
atmospheric Cherenkov telescopes located in southern Arizona~\cite{Weekes}. 
Designed to detect emission from astrophysical
objects in the energy range from 100~GeV to greater than 30~TeV,
VERITAS has an energy resolution of $\sim$15\% and an angular
resolution (68\% containment) of $\sim$0.1$^\circ$ per event at 1~TeV.  
A source with a flux of 1\% of the Crab Nebula flux is detected in $\sim$25~hours
of observations, while a \mbox{5\% Crab Nebula} flux source is detected in less than
2~hours.  The field of view of the VERITAS telescopes is
3.5$^\circ$.  For more details on the VERITAS instrument and the imaging atmospheric-Cherenkov
technique, see~\cite{Perkins2009}.

\section{Observations}

Most of the VERITAS observations of radio galaxies are on the radio galaxy M~87.
This AGN is located in the center of the Virgo cluster at a distance of $\sim$16~Mpc
and is currently the brightest detected VHE radio galaxy.
M~87 was originally detected with marginal significance by HEGRA at TeV 
energies~\cite{Aharonian2003}, 
and later also by HESS~\cite{Aharonian2006}, VERITAS~\cite{Acciari2008} and MAGIC~\cite{Albert2008}.
This giant radio galaxy has always been of particular interest because
its jet lies at $\sim$20$^\circ$ respect to the line of sight and
its core and the structure of the jet
are spatially resolved in X-ray, optical and radio observations,
thus it is an ideal candidate for correlated MWL studies~\cite{Wilson2002}.

In 2008 VERITAS coordinated an observational campaign with two other major VHE observatories
(MAGIC, HESS), overlapping with VLBA radio observations~\cite{M87Science}. 
Three Chandra X-ray pointed observations have also been performed during the first half of 2008.
Multiple flares at VHE have been detected. In X-rays, the inner-most knot in the jet (HST-1) was found in low
state, while the core region was in high state since 2000. Progressive brightening of the core region
in radio was also seen along the VHE flare development. This is an indication that 
the $\gamma$-ray emission originates from a region close to the core rather than from more distant regions.

\begin{figure}[!t]
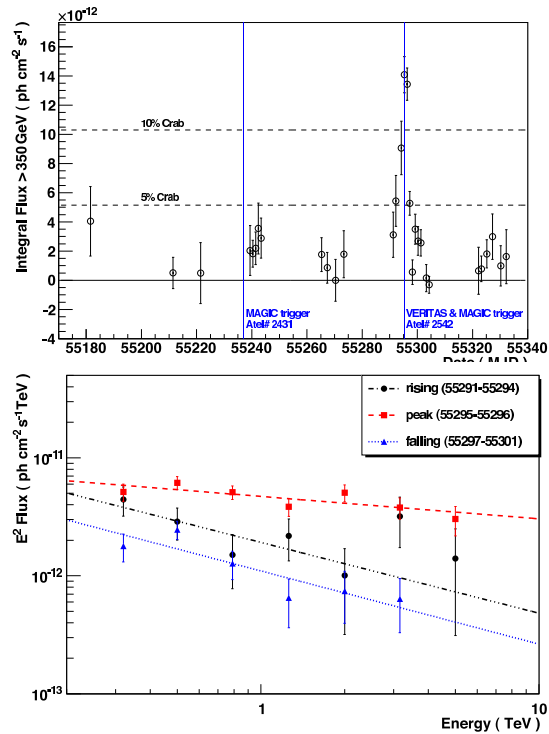

  \vspace{5mm}
  \centering
  \includegraphics[width=2.8in]{icrc0781_fig02.epsi}
  \includegraphics[width=2.8in]{icrc0781_fig03.epsi}
  \caption{\emph{(upper plot)} VERITAS light curve of the 2010 seasonal monitoring campaign.
  \emph{(lower plot)} Spectral analysis for three phases of the April 2010 flaring event: rising phase ( circles),
  peak (squares) and decreasing phase (triangles).}
  \label{fig_M87-2}
 \end{figure}

In April 2010, during the seasonal monitoring of M~87, VERITAS detected another flare with peak flux 
of $\sim$20\% of the Crab Nebula flux. During the six-month observation period, M~87 was detected 
at a level of 25.6$\sigma$ above the background, with an average flux above 350~GeV equivalent 
to 5\% of the Crab Nebula flux. Dedicated analysis in 20-minute bins has been performed on the 
April 2010 flaring episode. A spectral analysis has been done on three different phases of the
flaring episode: the rising phase, the peak and the falling phase. A power-law fit has been applied to each phase,
showing a hint of spectral variability: $\Gamma_\mathrm{rise}=2.60\pm0.31$, 
$\Gamma_\mathrm{peak}=2.19\pm0.07$, $\Gamma_\mathrm{fall}=2.62\pm0.18$. Figure~\ref{fig_M87-2}
shows the 2010 seasonal lightcurve and the spectral analysis at different times for the flaring episode 
that occurred in April 2010. Details on the analysis and results of the 2010 observational campaign on M~87
are presented in a publication currently in the process of peer-review. Results of the extensive multi-year
MWL observational campaign on M~87 will be presented soon too.

\section{Conclusions}

The radio galaxy M~87 is a unique laboratory for studying the acceleration and emission processes
around the supermassive black hole of AGNs. Its relatively high brightness in VHE $\gamma$-rays
enables to perform cross-correlated MWL observational campaigns and to study variability and spectral evolution
features. VERITAS VHE observations have been crucial during past MWL observational campaigns in
identifying a close-region to the core as responsible for the $\gamma$-ray emission. During the 2009-2010 observational
season, VERITAS detected the strongest flare ever observed in $\gamma$-rays on M~87.
This observation enabled for the first time the study of flux and spectral temporal properties on a radio galaxy.
Further details on the long-term MWL observational campaign on M~87 are in the process of publication.
\bigskip 
\begin{acknowledgments}

This research is supported by grants from the US Department of Energy, the US National Science Foundation, 
and the Smithsonian Institution, by NSERC in Canada, by Science Foundation Ireland, and by STFC in the UK. 
We acknowledge the excellent work of the technical support staff at the FLWO and at the collaborating 
institutions in the construction and operation of the instrument.

\end{acknowledgments}

\bigskip 

\begin{thebibliography}{9}   
\bibitem{Aharonian2003} Aharonian, F., et al. 2003, \emph{A\&A}, \textbf{403}, L1
\bibitem{Acciari2008} Acciari, V. A., et al. 2008, \emph{ApJ}, \textbf{679}, 397
\bibitem{Aharonian2009} Aharonian F. 2009, \emph{ApJ}, \textbf{695}, L40
\bibitem[Fanaroff \& Riley(1974)]{FanaroffRiley} Fanaroff, B. F., Riley, J. M. 1974, \emph{MNRAS}, \textbf{167}, 31
\bibitem[Zirbel \& Baum(1995)]{Zirbel1995} Zirbel, E. L., Baum S. A. 1995, \emph{ApJ}, \textbf{448}, 521
\bibitem[Urry \& Padovani(1995)]{UrryPadovani}Urry, C. M., Padovani, P. 1995, \emph{PASP}, \textbf{54}, 215
\bibitem[Ghisellini et al.(1993)]{Ghisellini1993} Ghisellini, G., et al. 1993, \emph{ApJ}, \textbf{407}, 65
\bibitem[Jones et al.(1974)]{Jones1974} Jones, T. W., O'Dell, S. L., Stein, W. A. 1974, \emph{ApJ}, \textbf{188}, 353
\bibitem[Bloom \& Marscher(1996)]{BloomMarscher1996} Bloom, S. D., Marscher, A. P. 1996 \emph{ApJ}, \textbf{461}, 657
\bibitem[Ghisellini et al.(1989)]{Ghisellini1989} Ghisellini, G., George, I. M., Done, C. 1989, \emph{MNRAS}, \textbf{241}, 43
\bibitem[Mannheim(1993)]{Mannheim1993} Mannheim, K. 1993, \emph{A\&A}, \textbf{269}, 67
\bibitem[Aharonian(2002)]{Aharonian2002} Aharonian, F. 2002, \emph{MNRAS}, \textbf{332}, 215
\bibitem[Reimer et al.(2004)]{Reimer2004} Reimer, A., et al. 2004, \emph{A\&A}, \textbf{419}, 89
\bibitem[Beall \& Bednarek(1999)]{Beall1999} Beall J.H., Bednarek W. 1999, \emph{ApJ}, \textbf{510}, 188
\bibitem[Pohl \& Schlickeiser(2000)]{Pohl2000} Pohl M., Schlickeiser R. (2000), \emph{A\&A}, \textbf{354}, 395
\bibitem[Krawczynski et al.(2002)]{Kraw2002} Krawczynski, H., Coppi, P. S., \& Aharonian., F. 2002,
\emph{MNRAS}, \textbf{336}, 721
\bibitem[Marscher(1999)]{Marscher1999} Marscher, A. P. 1999, \emph{Astropart. Phys.}, \textbf{11}, 19
\bibitem[Georganopoulos \& Kazanas(2003)]{Georganopoulos2003} Georganopoulos, M., \& Kazanas, D. 2003
\emph{ApJ}, \textbf{589}, L5
\bibitem[Ghisellini et al.(2005)]{Ghisellini2005} Ghisellini, G., et al. 2005, \emph{A\&A}, \textbf{432}, 401
\bibitem{Weekes} Weekes, T.~C., et~al., Astroparticle Physics, 2002, {\bf 17}, 221-243
\bibitem{Perkins2009} Perkins, J. S., eConf Proceedings C091122, astro-ph:0912.3841
\bibitem{Aharonian2006} Aharonian, F., et al., Science, 2006, {\bf 314}, 1424-1427
\bibitem{Albert2008} Albert, J., et al., ApJ, 2008, {\bf 685}, L23-L26
\bibitem{Wilson2002} Wilson, A. S., \& Yang, Y., ApJ, 2002, {\bf 568}, 133-140
\bibitem{M87Science} Acciari, V. A., et al., Science, 2009, {\bf 325}, 444




\end{thebibliography}

\end{document}